\documentclass[aps,prd,nofootinbib,superscriptaddress]{revtex4}
\usepackage{graphicx}
\usepackage{dcolumn}
\usepackage{amsmath}
\usepackage{epsfig}
\usepackage{colordvi}
\usepackage{color}
\usepackage{hhline}

\begin{document}
\title{\large \bf \boldmath
Updated measurement of the $e^+e^- \to \omega \pi^0 \to
\pi^0\pi^0\gamma$ cross section with the SND detector}

\author{M.~N.~Achasov}
\affiliation{Budker Institute of Nuclear Physics, SB RAS, Novosibirsk, 630090, Russia}
\affiliation{Novosibirsk State University, Novosibirsk, 630090, Russia}
\author{A.~Yu.~Barnyakov}
\author{K.~I.~Beloborodov}
\author{A.~V.~Berdyugin}
\author{D.~E.~Berkaev}
\affiliation{Budker Institute of Nuclear Physics, SB RAS, Novosibirsk, 630090, Russia}
\affiliation{Novosibirsk State University, Novosibirsk, 630090, Russia}
\author{A.~G.~Bogdanchikov}
\author{A.~A.~Botov}
\affiliation{Budker Institute of Nuclear Physics, SB RAS, Novosibirsk, 630090, Russia}
\author{T.~V.~Dimova}
\affiliation{Budker Institute of Nuclear Physics, SB RAS, Novosibirsk, 630090, Russia}
\affiliation{Novosibirsk State University, Novosibirsk, 630090, Russia}
\author{V.~P.~Druzhinin}
\author{V.~B.~Golubev}
\author{L.~V.~Kardapoltsev}
\email[e-mail: ]{l.v.kardapoltsev@inp.nsk.su}
\affiliation{Budker Institute of Nuclear Physics, SB RAS, Novosibirsk, 630090, Russia}
\affiliation{Novosibirsk State University, Novosibirsk, 630090, Russia}
\author{A.~S.~Kasaev}
\affiliation{Budker Institute of Nuclear Physics, SB RAS, Novosibirsk, 630090, Russia}
\author{A.~G.~Kharlamov}
\affiliation{Budker Institute of Nuclear Physics, SB RAS, Novosibirsk, 630090, Russia}
\affiliation{Novosibirsk State University, Novosibirsk, 630090, Russia}
\author{A.~N.~Kirpotin}
\affiliation{Budker Institute of Nuclear Physics, SB RAS, Novosibirsk, 630090, Russia}
\author{D.~P.~Kovrizhin}
\affiliation{Budker Institute of Nuclear Physics, SB RAS, Novosibirsk, 630090, Russia}
\affiliation{Novosibirsk State University, Novosibirsk, 630090, Russia}
\author{I.~A.~Koop}
\affiliation{Budker Institute of Nuclear Physics, SB RAS, Novosibirsk, 630090, Russia}
\affiliation{Novosibirsk State University, Novosibirsk, 630090, Russia}
\affiliation{Novosibirsk State Technical University, Novosibirsk, 630092,Russia}
\author{A.~A.~Korol}
\affiliation{Budker Institute of Nuclear Physics, SB RAS, Novosibirsk, 630090, Russia}
\affiliation{Novosibirsk State University, Novosibirsk, 630090, Russia}
\author{S.~V.~Koshuba}
\affiliation{Budker Institute of Nuclear Physics, SB RAS, Novosibirsk, 630090, Russia}
\author{K.~A.~Martin}
\affiliation{Budker Institute of Nuclear Physics, SB RAS, Novosibirsk, 630090, Russia}
\author{A.~E.~Obrazovsky}
\affiliation{Budker Institute of Nuclear Physics, SB RAS, Novosibirsk, 630090, Russia}
\author{E.~V.~Pakhtusova}
\affiliation{Budker Institute of Nuclear Physics, SB RAS, Novosibirsk, 630090, Russia}
\author{Yu.~A.~Rogovsky}
\affiliation{Budker Institute of Nuclear Physics, SB RAS, Novosibirsk, 630090, Russia}
\affiliation{Novosibirsk State University, Novosibirsk, 630090, Russia}
\author{A.~I.~Senchenko}
\affiliation{Budker Institute of Nuclear Physics, SB RAS, Novosibirsk, 630090, Russia}
\author{S.~I.~Serednyakov}
\author{Z.~K.~Silagadze}
\author{Yu.~M.~Shatunov}
\author{D.~A.~Shtol}
\author{D.~B.~Shwartz}
\affiliation{Budker Institute of Nuclear Physics, SB RAS, Novosibirsk, 630090, Russia}
\affiliation{Novosibirsk State University, Novosibirsk, 630090, Russia}
\author{A.~N.~Skrinsky}
\affiliation{Budker Institute of Nuclear Physics, SB RAS, Novosibirsk, 630090, Russia}
\author{I.~K.~Surin}
\author{A.~V.~Vasiljev}
\affiliation{Budker Institute of Nuclear Physics, SB RAS, Novosibirsk, 630090, Russia}
\affiliation{Novosibirsk State University, Novosibirsk, 630090, Russia}

\begin{abstract}
We analyze a 37 pb$^{-1}$ data sample collected with the SND detector at 
the VEPP-2000 $e^+e^-$ collider in the center-of-mass energy range
1.05--2.00 GeV and present an updated measurement of the $e^+e^- \to \omega
\pi^0 \to \pi^0\pi^0\gamma$ cross section. In particular, we correct
the mistake in radiative correction calculation made in our previous
measurement based on a part of the data. The measured cross section is 
fitted with the vector meson dominance model with three $\rho$-like states 
and used to test the conserved vector current hypothesis in the 
$\tau^-\to\omega\pi^-\nu_{\tau}$ decay.
\end{abstract}

\maketitle

\section{Introduction}
The process
$e^+e^- \to \omega \pi^0$ is one of the dominant 
processes contributing to the total hadronic cross section at the
center-of-mass (c.m.) energy between 1 and 2 GeV. The measurement of
the $e^+e^- \to \omega \pi^0$ cross section provides
information about properties of excited $\rho$-like states and
can be used to check the relation between the cross section and
the spectral function in the $\tau\to\omega\pi^-\nu_{\tau}$ decay
following from the conserved vector current hypothesis~\cite{Tsai}.

The $e^+e^- \to \omega \pi^0$ cross section was measured in the 
ND~\cite{Dolinsky}, SND~\cite{ppg_SND,ppg_SND_phi,4pi_snd}, 
and CMD-2~\cite{4pi_cmd,ppg_CMD} experiments
at the VEPP-2M collider at c.m. energies below 1.4 GeV, in
the KLOE experiment~\cite{KLOE} in the $\phi$-meson region, and
in the DM2 experiment~\cite{dm2} in the $1.35-2.4$ GeV energy range.

In the SND experiment~\cite{snd1,snd2,snd3,snd4} at the VEPP-2000 $e^+e-$ 
collider~\cite{vepp2k} the $e^+e^- \to \omega \pi^0$ cross section
is studied in the five-photon final state when the of $\omega$ meson decays
to $\pi^0\gamma$. The result of the study based on data collected in 2010
and 2011 was published in Ref.~\cite{ompi2013}. However, a mistake was made 
in the calculation of the radiative corrections in Ref.~\cite{ompi2013},
which led to incorrect measurement of the Born $e^+e^- \to \omega \pi^0$ cross
section. 

In this paper we reanalyze the 2010-2011 data sample (25 pb$^{-1}$) and add 
data
collected in 2012 (12 pb$^{-1}$). The analysis is very close to that described
in Ref.~\cite{ompi2013}. We correct the mistake in the radiative correction 
calculation and slightly modify the selection criteria of 
$e^+e^-\to\gamma\gamma$ events for luminosity measurement. The analysis uses
an updated version of the event reconstruction and simulation software.
Therefore, the values of the efficiency corrections and systematic 
uncertainties are changed compared with those in Ref.~\cite{ompi2013}.

In this analysis we use the corrected values of the c.m. energy ($E$) 
obtained in 
Ref.~\cite{KKpp}. The accuracy of the energy measurement is 6 MeV and 2 MeV
for 2011 and 2012 experiments, respectively. The 2010 and 2011 data  
are combined assuming that the energy calibration is the same
for the 2010 and 2011 experiments. 

\section{Luminosity measurement}
Following the previous work~\cite{ompi2013}, we use the 
$e^+e^- \to \gamma\gamma$ 
process for the luminosity measurement. Similar to the process under study,
it does not contain charged particles in the final state. Therefore,
some uncertainties on the cross section measurement cancel as a result of 
the luminosity normalization. 

We select events with at least two photons and no charged particles.
The number of hits in the detector drift chamber is required to not exceed 5.
The energies of two most energetic photons in an event must be
larger than $0.3E$. The azimuthal ($\phi_{1,2}$) and polar ($\theta_{1,2}$) 
angles of these photons must satisfy the following conditions:
$||\phi_1-\phi_2|-180^\circ| < 11.5^\circ$, 
$|\theta_1+\theta_2-180^\circ|< 17.2^\circ$, 
and $(180^{\circ}-|\theta_{1}-\theta_{2}|)/2 > 54^{\circ}$.

The integrated luminosity measured for each energy point is listed in 
Table~\ref{allres}. The systematic uncertainty on the luminosity measurement
estimated to be 1.4\%.

\section{Event selection}
For the process under study 
$e^+e^- \to \omega \pi^0\to \pi^0 \pi^0 \gamma \to 5\gamma$,
candidate events must have at least five photons and no charged particles.
The number of hits in the drift chamber is required to not exceed 5.
The normalized total energy deposition in the calorimeter 
$E_{\rm tot}/E > 0.5$. For events passing this selection, kinematic fits to 
the $e^+e^- \to 5\gamma$ and $e^+e^- \to \pi^0 \pi^0 \gamma $ hipotheses
are performed. The following cuts are applied on the $\chi^2$ of the
kinematic fits: $\chi^2_{5\gamma}<30$ and 
$\chi^2_{\pi^0\pi^0\gamma}-\chi^2_{5\gamma}<30$. The detailed description 
of the kinematic fits and the selection criteria can be found in 
Ref.~\cite{ompi2013}. 
\begin{figure}
\includegraphics[width=0.45\textwidth]{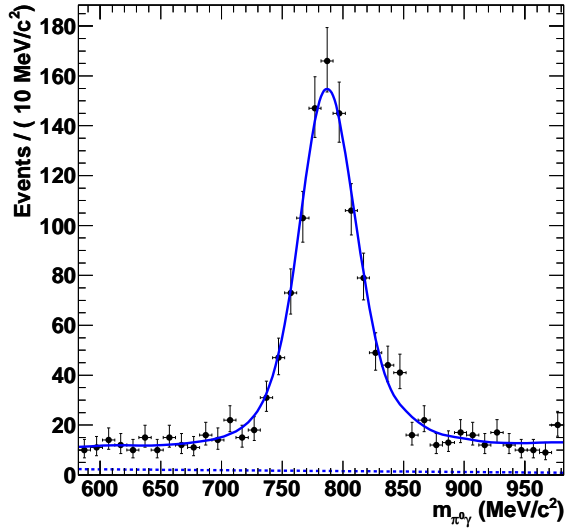}
\caption{ The distribution of the $\pi^0\gamma$ invariant mass
for selected data events (points with error bars) with 
$E=1494$ MeV.
The curve is the result of the fit by the sum of the signal and background
distribution. The dashed curve represents the background contribution.
\label{fig1}}
\end{figure}


The number of signal events is determined from the fit to the $\pi^0\gamma$
mass spectrum with a sum of signal and background distributions. The fitting
procedure is described in detail in Ref.~\cite{ompi2013}. The result of
the fit to the data spectrum for the energy point $E=1494$ MeV 
(two entries per event) is shown in Fig.~\ref{fig1}. 
The $\omega$ meson peak is clearly seen in the spectrum.
The dashed curve represents the background contribution. 
The fitted numbers of signal events 
for each energy point is listed in Table~\ref{allres}.
\begin{table}
\caption{The c.m.~energy ($E$), integrated luminosity ($L$), detection
efficiency ($\varepsilon$), number of selected signal events ($N_{s}$),
radiative-correction factor ($1+\delta$), measured Born cross section
($\sigma$). The quoted errors on $N_{s}$ and $\sigma$ are statistical.
The systematic uncertainty on the cross section is 2.7\% at $E<1.59$ GeV,
3.5\% at $1.59 < E <1.79$~GeV
and 5.2\% in energy range $E>1.79$ GeV.
\label{allres}}
\begin{ruledtabular}
\begin{tabular}{cccccc}
 $E$, GeV & $L$, nb$^{-1}$ & $\varepsilon$, \% & $N_{s}$ & $1+\delta$ &
 $\sigma$, nb \\[0.3ex] \hline \\[-2.1ex]
$1047 \pm 6$  & 352  & 35.1  &  $105\pm 13$ &0.895  & $0.95 \pm  0.11$             \\                       
$1075 \pm 6$  & 542  & 35.3  &  $180\pm 17$ &0.903  & $1.04 \pm  0.09$             \\                       
$1097 \pm 6$  & 839  & 35.5  &  $290\pm 19$ &0.905  & $1.08 \pm  0.07$             \\                       
$1124 \pm 6$  & 520  & 35.9  &  $213\pm 17$ &0.907  & $1.26 \pm  0.11$             \\                       
$1151 \pm 6$  & 412  & 36.3  &  $178\pm 13$ &0.912  & $1.3 \pm  0.1  $               \\                       
$1174 \pm 6$  & 536  & 36.3  &  $230\pm 21$ &0.913  & $1.29 \pm  0.12$              \\                       
$1196 \pm 6$  & 1063  & 36.0 &  $485\pm 25$ &0.913  &$1.39 \pm  0.07$             \\                       
$1223 \pm 6$  & 554  & 37.2  &  $251\pm 19$ &0.913  & $1.33 \pm  0.11$             \\                       
$1245 \pm 6$  & 432  & 37.3  &  $184\pm 14$ &0.913  & $1.25 \pm  0.12$             \\                       
$1273 \pm 6$  & 495  & 37.1  &  $257\pm 21$ &0.914  & $1.53 \pm  0.13$             \\                       
$1277 \pm 2$  & 677  & 37.3  &  $320\pm 21$ &0.917  & $1.38 \pm  0.09$             \\                       
$1295 \pm 6$  & 1266  & 37.5 &  $678\pm 31$ &0.915  &$1.56 \pm  0.07$             \\                       
$1323 \pm 6$  & 526  & 38.2  &  $282\pm 23$ &0.915  & $1.54 \pm  0.12$             \\                       
$1344 \pm 6$  & 553  & 37.8  &  $289\pm 24$ &0.917  & $1.5 \pm  0.13 $              \\                       
$1357 \pm 2$  & 756  & 37.8  &  $418\pm 29$ &0.915  & $1.6 \pm  0.11 $             \\                       
$1374 \pm 6$  & 572  & 37.5  &  $304\pm 23$ &0.916  & $1.55 \pm  0.12$             \\                       
$1394 \pm 6$  & 1042  & 37.3 &  $574\pm 32$ &0.921  &$1.61 \pm  0.09$              \\                       
$1423 \pm 6$  & 598  & 37.9  &  $372\pm 22$ &0.922  & $1.78 \pm  0.1 $              \\                       
$1435 \pm 2$  & 917  & 37.4  &  $528\pm 27$ &0.923  & $1.67 \pm  0.09$             \\                       
$1443 \pm 6$  & 428  & 37.4  &  $218\pm 17$ &0.926  & $1.47 \pm  0.12$             \\                       
$1471 \pm 6$  & 596  & 37.6  &  $285\pm 20$ &0.931  & $1.37 \pm  0.1 $              \\                       
$1494 \pm 6$  & 1954  & 38.1 &  $990\pm 40$ &0.938  &$1.42 \pm  0.06$             \\                       
$1515 \pm 2$  & 829  & 37.6  &  $355\pm 22$ &0.944  & $1.21 \pm  0.08$             \\                       
$1522 \pm 6$  & 478  & 37.6  &  $251\pm 17$ &0.945  & $1.48 \pm  0.1 $              \\                       
$1543 \pm 6$  & 546  & 38.2  &  $225\pm 16$ &0.952  & $1.13 \pm  0.08$             \\                       
$1572 \pm 6$  & 510  & 37.4  &  $170\pm 15$ &0.965  & $0.93 \pm  0.08$             \\                       
$1595 \pm 2$  & 903  & 37.3  &  $264\pm 19$ &0.976  & $0.77 \pm  0.07$             \\                       
$1594 \pm 6$  & 820  & 36.7  &  $226\pm 19$ &0.974  & $0.8 \pm  0.06 $             \\                       
$1623 \pm 6$  & 508  & 37.6  &  $132\pm 14$ &0.992  & $0.7 \pm  0.08 $             \\                       
$1643 \pm 6$  & 471  & 36.4  &  $94\pm 11$  &1.006  &  $0.54 \pm  0.07$             \\                       
$1669 \pm 6$  & 454  & 36.2  &  $75\pm 11$  &1.021  &  $0.45 \pm  0.07$             \\                       
$1674 \pm 2$  & 837  & 35.7  &  $127\pm 13$ &1.025  & $0.41 \pm  0.04$             \\                       
$1693 \pm 6$  & 827  & 35.1  &  $105\pm 13$ &1.043  & $0.35 \pm  0.04$             \\                       
$1716 \pm 2$  & 455  & 34.6  &  $50\pm 7$   &1.059  &   $0.3 \pm  0.05 $             \\                       
$1723 \pm 6$  & 507  & 35.5  &  $32\pm 7$   &1.060  &   $0.17 \pm  0.04$             \\                       
$1742 \pm 6$  & 509  & 34.4  &  $32\pm 8$   &1.077  &   $0.17 \pm  0.04$             \\                       
$1758 \pm 2$  & 797  & 33.8  &  $50\pm 10$  &1.081  &   $0.17 \pm  0.04$             \\                       
$1774 \pm 6$  & 530  & 34.1  &  $38\pm 7$   &1.072  &   $0.2 \pm  0.04 $              \\                       
$1798 \pm 2$  & 919  & 32.2  &  $59\pm 10$  &1.066  &  $0.14 \pm  0.03$             \\                       
$1793 \pm 6$  & 752  & 32.8  &  $38\pm 7$   &1.058  &   $0.19 \pm  0.03$             \\                       
$1826 \pm 6$  & 488  & 33.4  &  $16\pm 5$   &1.037  &   $0.09 \pm  0.03$              \\                       
$1840 \pm 2$  & 953  & 31.9  &  $38\pm 8$   &1.024  &   $0.12 \pm  0.03$             \\                       
$1849 \pm 6$  & 403  & 31.7  &  $9\pm 5$    &1.020  &    $0.07 \pm  0.04$             \\                       
$1871 \pm 6$  & 641  & 31.6  &  $20\pm 6$   &1.002  &   $0.1 \pm  0.03 $              \\                       
$1874 \pm 2$  & 835  & 30.7  &  $32\pm 6$   &1.004  &   $0.12 \pm  0.02$             \\                       
$1893 \pm 6$  & 579  & 31.4  &  $24\pm 6$   &0.993  &   $0.13 \pm  0.03$             \\                       
$1903 \pm 2$  & 867  & 29.9  &  $16\pm 5$   &0.989  &   $0.08 \pm  0.03$             \\                       
$1901 \pm 6$  & 559  & 29.6  &  $13\pm 5$   &0.986  &   $0.06 \pm  0.02$             \\                       
$1926 \pm 2$  & 614  & 29.6  &  $14\pm 5$   &0.978  &   $0.08 \pm  0.03$             \\                       
$1927 \pm 6$  & 562  & 29.4  &  $16\pm 4$   &0.979  &   $0.1 \pm  0.03 $             \\                       
$1945 \pm 2$  & 823  & 28.8  &  $15\pm 5$   &0.973  &   $0.06 \pm  0.02$             \\                       
$1953 \pm 6$  & 402  & 29.3  &  $4\pm 3$    &0.970  &    $0.03 \pm  0.03$             \\                       
$1967 \pm 2$  & 679  & 27.7  &  $12\pm 4$   &0.970  &   $0.06 \pm  0.03$             \\                       
$1978 \pm 6$  & 467  & 27.1  & $5.6^{+6.5 }_{-3.0}$ &0.970  &    $0.05^{+ 0.05 }_{- 0.02}$  \\                                      
$1989 \pm 2$  & 578  & 27.6  &  $10\pm 3$   &0.964  &    $0.06 \pm  0.02$             \\                       
$2005 \pm 6$  & 546  & 27.2  &  $10\pm 4$   &0.965  &    $0.07 \pm  0.03$             \\  \hline 
\end{tabular}	  
\end{ruledtabular}
\end{table}

\section{\boldmath The $e^+e^- \to \omega \pi^0\to \pi^0 \pi^0 \gamma$ Born cross 
section}
The experimental values of the Born cross section is determined as
\begin{equation}
\sigma(E_i)=\frac{N_{s,i}}{L_i\varepsilon_i(1+\delta(E_i))},
\label{born}
\end{equation}
where $N_{s,i}$, $L_i$, $\varepsilon_i$, and $\delta(E_i)$ are 
the number of selected signal events, integrated luminosity, 
detection efficiency, and radiative correction for the $i$-th energy point.
The detection efficiency for the process under study is determined
using MC simulation and then corrected by $(-3.9\pm0.7)\%$ to take into account
data-MC simulation difference in the detector response
(see Ref.~\cite{ompi2013} for details). The found detection 
efficiency $\varepsilon_r$ is a function of two parameters: the c.m.~energy 
$E$ and the energy of the extra photon $E_r$ emitted from the initial state. 
The efficiency in Eq.~(\ref{born}) $\varepsilon_i=\varepsilon_r(E_i,E_r=0)$.

The radiative correction is determined from the fit to the visible cross
section data ($\sigma_{\rm vis}(E_i)=N_{s,i}/L_i$) with the function
\begin{equation}
\sigma_{vis}(E) = \sigma(E)\varepsilon(E)(1+\delta(E)=
\int \limits_{0}^{x_{max}}\varepsilon_r(E, xE/2)F(x,E)\sigma(E\sqrt{1-x})dx,
\label{viscrs}
\end{equation}
where $F(x,E)$ is a function describing the probability to emit extra photons 
with the total energy $xE/2$ from the initial state~\cite{FadinRad}.
The Born cross section
is described by the following formula~\cite{ompi2013}
\begin{equation}
\sigma(E)=\frac{4\pi\alpha^2}{3E^3}
|F_{\omega\pi\gamma}(E^2)|^2P_f(E)B(\omega\to\pi^0\gamma),
\label{FormEq1}
\end{equation}
where $\alpha$ is the fine structure constant, $F_{\omega\pi\gamma}(E^2)$ is
the $\gamma^\ast\to \omega\pi^0$ transition form factor, 
$B(\omega\to\pi^0\gamma)$ is the branching fraction for the 
$\omega\to\pi^0\gamma$ decay, and $P_f(E)$ is
the phase-space factor. In the narrow $\omega$-resonance approximation
$P_f(E)=q_{\omega}^3$, where $q_{\omega}$ is the $\omega$-meson 
momentum. The transition form factor is parametrized in the vector
meson dominance (VMD) model as a sum of the
$\rho(770)$, $\rho(1450)$, and $\rho(1700)$ resonance contributions:
\begin{equation}
\label{FormEq2} F_{\omega\pi\gamma}(E^2) =
\frac{g_{\rho\omega\pi}}{f_{\rho}}
 \left(\frac{m_{\rho}^2}{D_{\rho}}+
 A_1 e^{i\varphi_1}\frac{m_{\rho(1450)}^2}{D_{\rho(1450)}}+ 
 A_2 e^{i\varphi_2}\frac{m_{\rho(1700)}^2}{D_{\rho(1700)}} \right),
\end{equation}
where $g_{\rho\omega\pi}$ is the $\rho\to \omega\pi$ coupling constant,
$f_{\rho}$ is the $\gamma^{*}\to \rho$ coupling constant calculated from
the $\rho \to e^+e^-$ decay width,
$D_{\rho_i}(E) = m_{\rho_i}^2-E^2-\imath E \Gamma_{\rho_i}(E)$,
$m_{\rho_i}$ and $\Gamma_{\rho_i}(E)$ are the mass and width of the
resonance $\rho_i$. The formula for $\Gamma_{\rho(770)}(E)$ is given in
Ref.~\cite{ompi2013}. For the $\rho(1450)$ and
$\rho(1700)$, the energy-independent widths are used.

The free fit parameters are
$g_{\rho\omega\pi}$, $m_{\rho(1450)}$, $\Gamma_{\rho(1450)}$, 
and the relative amplitudes ($A_i$) and phases ($\varphi_i$) for 
the $\rho(1450)$ and $\rho(1700)$. Since the  $\rho(1700)$ contribution is 
found to be small, its mass and width are fixed at their Particle Data Group 
(PDG) values~\cite{pdg}. 
The data from this work are fitted together with the cross-section
data obtained in the SND@VEPP-2M~\cite{ppg_SND,ppg_SND_phi} and 
KLOE~\cite{KLOE} experiments. The model describes data well 
($\chi^2/\nu=71/73$, where $\nu$ is the number of degrees of freedom). 
The obtained fit parameters listed in Table~\ref{fitres} (Model 1)   
are used to calculate the radiative corrections.
\begin{table}
\caption{The fitted parameters of the $e^+e^- \to \omega \pi^0$
cross-section model.\label{fitres}}
\begin{ruledtabular}
\begin{tabular}{lccc}
 Parameter & Model 1 & Model 2 & Model 3\\ [0.3ex] \hline \\[-2.1ex]
 $g_{\rho\omega\pi}$, GeV$^{-1}$ & 15.9 $\pm$ 0.4 & 16.5 $\pm$ 0.2 & --\\
$A_1$ & 0.175 $\pm$ 0.016 & 0.137 $\pm$ 0.006 & 0.251 $\pm$ 0.006\\
 $A_2$ & 0.014 $\pm$ 0.004 & $\equiv$ 0 & 0.027 $\pm$ 0.003\\
 $M_{\rho(1450)}$, MeV & 1510 $\pm$ 7 & 1499 $\pm$ 4 & 1516 $\pm$ 10\\
$\Gamma_{\rho(1450)}$, MeV & 440 $\pm$ 40 & 367 $\pm$ 13 & 500 $\pm$ 30\\
$ M_{\rho(1700)}$, MeV & $\equiv$ 1720 & -- & $\equiv$ 1720\\
$ \Gamma_{\rho(1700)}$, MeV & $\equiv$ 250 & -- & $\equiv$ 250\\
 $\varphi_1$, deg. & 124 $\pm$ 17 & 122 $\pm$ 8 & 162 $\pm$ 6\\
$\varphi_2$, deg. & -63 $\pm$ 21& -- & -24 $\pm$ 10\\
 $\chi^2/\nu$ & 71 / 73& 85 / 75 & 83 / 74\\
\end{tabular}
\end{ruledtabular}
\end{table}

The experimental values of the Born cross section obtained using 
Eq.~(\ref{born}) are listed in Table~\ref{allres} together with the
values of the detection efficiency and radiative correction.
The quoted errors are statistical. The systematic uncertainties 
(see Ref.~\cite{ompi2013} for details)
are summarized in Table~\ref{syserr} for three energy intervals.
\begin{table*}
\caption{The systematic uncertainties (\%) on the measured cross section
from different. The total uncertainty is the sum of all the contributions 
in quadrature.\label{syserr}}
\begin{ruledtabular}
\begin{tabular}{lccc}
 Source  & $E<1.59$ GeV & $1.59<E<1.79$ GeV & $E>1.79$ GeV  \\[0.3ex]
 \hline \\[-2.1ex]
Luminosity & 1.4  & 1.4 & 1.4    \\
Selection criteria  & 0.5 & 0.5 & 0.5\\
Beam background& 0.5 & 0.5 & 0.5 \\
Radiative correction  & 1 & 1 & 3 \\
Interference with $\rho^0\pi^0$  & 2 & 3 & 4 \\
\hline 
Total & 2.7 & 3.5 & 5.2 \\
\end{tabular}
\end{ruledtabular}
\end{table*}
Compared with Ref.~\cite{ompi2013} there is no the systematic uncertainty
associated with data-MC simulation difference in photon conversion.
In this analysis, the photon conversion probability before the drift
chamber is measured with high accuracy using $e^+e^-\to\gamma\gamma$ events.

\section{Discussion}
\begin{figure}
\includegraphics[width=0.7\textwidth]{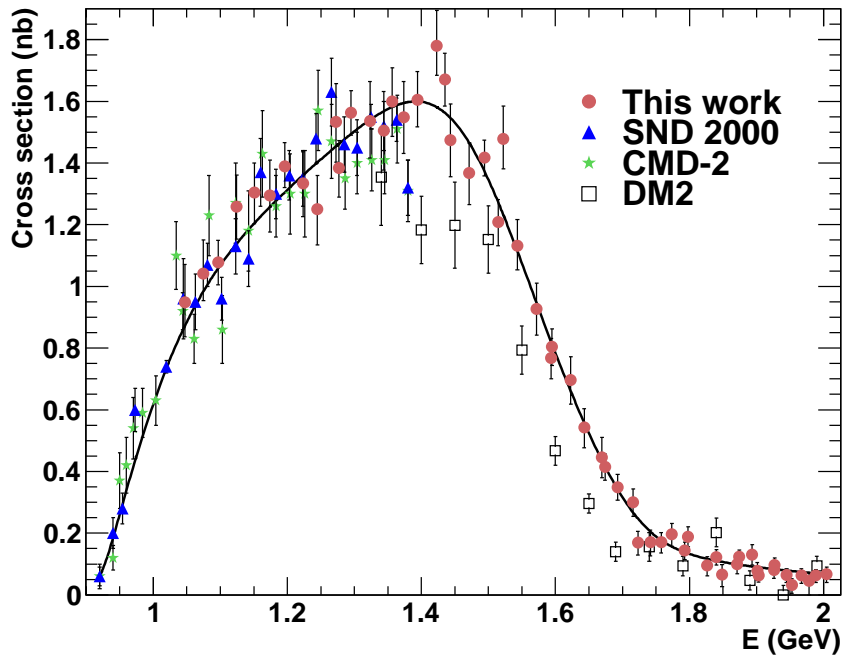}
\caption{The cross section for $e^+e^- \to \omega \pi^0 \to
\pi^0\pi^0\gamma$ measured in this work (circles), and in
SND@VEPP-2M~\cite{ppg_SND,ppg_SND_phi} (triangles),
CMD-2~\cite{ppg_CMD} (stars), and DM2~\cite{dm2} (squares) experiments.
Only statistical errors are shown.
The curve is the result of the fit to SND~2000 and SND~2013
data described in the text (Model 1).
\label{crs_ompi}}
\end{figure}
The measured $e^+e^-\to \omega\pi^0\to \pi^0\pi^0\gamma$ Born cross section 
is shown in Fig.~\ref{crs_ompi} in comparison with the previous 
measurements~\cite{ppg_SND,ppg_SND_phi,ppg_CMD,dm2}. Our data are in
agreement with the SND@VEPP-2M and CMD-2 measurements below 1.4 GeV, but exceed
the DM2 data in the energy region $1.35-1.75$ GeV.
After correction of the mistake made in Ref.~\cite{ompi2013},
the Born cross section increases by 2\% at $E=1.1$ GeV, 6\% at 1.4 GeV,
12\% at 1.6 GeV. Dramatic changes are observed above 1.8 GeV, where 
the cross section in Ref.~\cite{ompi2013} was consistent with zero.

The measured cross section is well described by the VDM model with
two excited $\rho$-like resonance. The fitted mass and width of the
$\rho(1450)$ resonance listed in Table~\ref{fitres} are in reasonable
agreement with the corresponding PDG values. The contribution of the
$\rho(1700)$ resonance is small. We also perform a fit with one
excited resonance (Model 2 in Table~\ref{fitres}). From the difference of 
the $\chi^2$ values for Models 1 and 2 we determine that the significance
of the $\rho(1700)$ contribution is $3.7\sigma$.

Using the fit results and the branching fraction 
$B(\omega\to\pi^0\gamma) = (8.88 \pm 0.18)\%$ measured by SND~\cite{SNDpg}, 
the products of the the branching fractions are calculated to be
\begin{eqnarray}
B(\rho(1450)\to e^+e^-)B(\rho(1450)\to\omega\pi^0)& = &
(2.1 \pm 0.4)\times 10^{-6},\\
B(\rho(1700)\to e^+e^-)B(\rho(1700)\to\omega\pi^0)& = &
(0.09 \pm 0.05)\times 10^{-6}.
\end{eqnarray}

\begin{figure}
\includegraphics[width=0.7\textwidth]{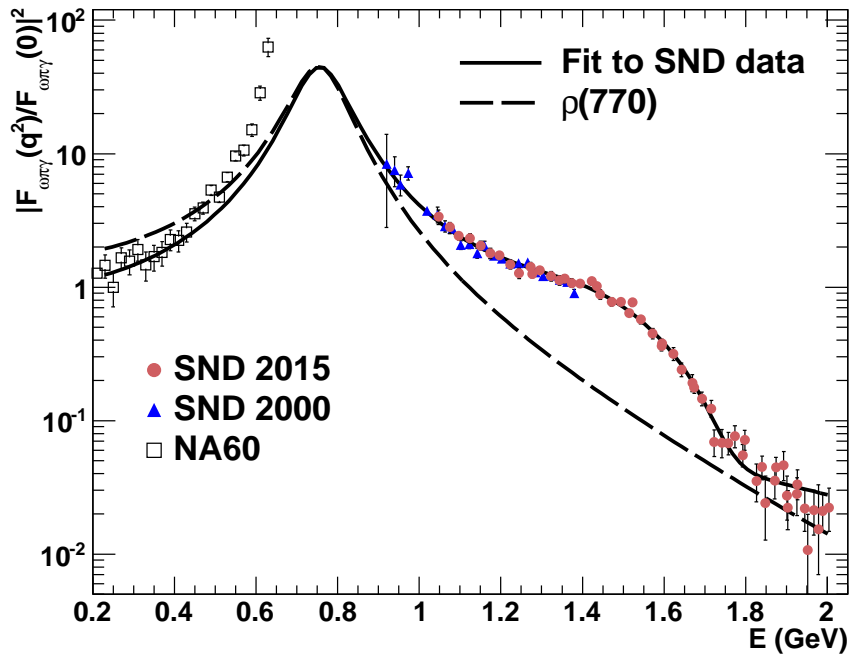}
\caption{The $\gamma^\ast\to\omega\pi^0$ transition form factor.
The points with error bars represent data from this work (circles),
Ref.~\cite{ppg_SND} (triangles), and Ref.~\cite{NA60} (squares).
Only statistical errors are shown.
The curve is the result of of the fit to the $e^+e^-\to\omega\pi^0$ 
cross-section data.  The dashed curve shows the
$\rho(770)$ contribution.\label{form}}
\end{figure}

In Fig.~\ref{form} we show our result on the normalized $\gamma^\ast\to \omega\pi^0$
transition form factor in comparison with the data obtained
from the $\omega\to \pi^0\mu^+\mu^-$ decay in the NA60 experiment~\cite{NA60}.
The $F_{\omega\pi\gamma}(0)$ value is calculated
from the $\omega\to\pi\gamma$ decay width measured by SND~\cite{ppg_SND}
using the formula
\begin{equation}
| F_{\omega\pi\gamma}(0)|^2=
\frac{3\Gamma(\omega\to\pi^0\gamma)}{\alpha P_\gamma^3},
\end{equation}
where $P_\gamma$ is the decay photon momentum.
We modify our form-factor model to provide correct normalization at zero:
\begin{equation}
\label{FormEq3} F_{\omega\pi\gamma}(E^2) =
F_{\omega\pi\gamma}(0)
 \left(\frac{m_{\rho}^2}{D_{\rho}}+
 A_1 e^{i\varphi_1}\frac{m_{\rho(1450)}^2}{D_{\rho(1450)}}+ 
 A_2 e^{i\varphi_2}\frac{m_{\rho(1700)}^2}{D_{\rho(1700)}} \right)/
 \left(1 + A_1 e^{i\varphi_1} + A_2 e^{i\varphi_2} \right).
\end{equation}
The parameters of this model obtained from the fit to
the SND and KLOE data are listed in Table~\ref{fitres} (Model 3).
The normalization requirement leads to worse but still acceptable 
fit quality. The fitted curve is shown in Fig.~\ref{form}.  It is seen that
the VMD model cannot describe simultaneously the 
$e^+e^-\to\omega\pi^0$ and $\omega\to \pi^0\mu^+\mu^-$ data.

The CVC hypothesis and isospin symmetry give the relation between 
the $e^+e^-\to\omega\pi^0$ cross section and the 
$\tau^-\to\omega\pi^-\nu_{\tau}$ decay width~\cite{Tsai,cleo}
\begin{equation}
\label{tauBr}
\Gamma(\tau^-\to\omega\pi^-\nu_{\tau})=
\frac{G_F^2 |V_{ud}|^2}{64\pi^4 \alpha^2
m_{\tau}^3}\int^{m_{\tau}}q^3(m_{\tau}^2-q^2)^2(m_{\tau}^2+2q^2)\sigma_{\omega\pi^0}(q)dq,
\end{equation}
where $|V_{ud}|$ is the Cabibbo-Kobayashi-Maskawa matrix element,
$m_{\tau}$ is the $\tau$ lepton mass, $G_F$ is the Fermi constant.
Integrating the fitted curve shown in Fig.~\ref{crs_ompi} we obtain
$\Gamma(\tau^-\to\omega\pi^-\nu_{\tau})B(\omega\to\pi^0\gamma)=
(3.76\pm0.04\pm0.10 )\times 10^{-6}$ eV.  
Using the PDG value of the $\tau$ lifetime~\cite{pdg} and the SND result for
$B(\omega\to\pi^0\gamma)$~\cite{SNDpg} we calculate
\begin{equation}
\label{tauBr2} B(\tau^-\to\omega\pi^-\nu_{\tau}) = (1.87 \pm 0.02 \pm 0.07)
\times 10^{-2}.
\end{equation}
The calculated branching fraction is in good agreement with the experimental 
value $(1.96 \pm 0.08) \times 10^{-2}$ that was obtained as a difference of
the PDG~\cite{pdg} values for $B(\tau^-\to\omega h^-\nu_{\tau})$
and $B(\tau^-\to\omega K^-\nu_{\tau})$.

\section{Summary}
The $e^+e^-\to\omega\pi^0\to\pi^0\pi^0\gamma$ cross section has been
measured in the energy range of 1.05--2.00 GeV in the experiment 
with the SND detector at the VEPP-2000 $e^+e^-$ collider. 
We correct the mistake in the radiative correction calculation made
in our previous work~\cite{ompi2013} and increase the statistics.
The results presented in this paper correct and supersede the results of 
Ref.~\cite{ompi2013}. 

Our data on the $e^+e^-\to\omega\pi^0$ cross section in the 
energy range 1.4--2.0 GeV are most accurate to date. Below 1.4 GeV
they agree with the SND@VEPP-2M~\cite{ppg_SND} and CMD-2~\cite{ppg_CMD} 
measurements. Significant disagreement is observed with DM2 data~\cite{dm2} 
in the energy range 1.35--1.75 GeV.

Data on the $e^+e^-\to\omega\pi^0$ cross section are well fitted by
the VMD model with the $\rho(770)$, $\rho(1450)$, and $\rho(1700)$
resonances. However, this model cannot describe simultaneously the data on the
$\gamma^\ast\to\omega\pi^0$ transition form obtained from the 
$\omega\to\pi^0\mu^+\mu^-$ decay~\cite{NA60}.

We have also tested the CVC hypothesis calculating the branching fraction
for the $\tau^-\to\omega\pi^-\nu_{\tau}$ decay from our $e^+e^-\to \omega\pi$
data. The calculated branching fraction agrees with the measured value
within the experimental uncertainty of about 5\%.

\section{Acknowledgments}
This work is partly supported by the RFBR grant 15-02-01037.
Part of this work related to the photon reconstruction algorithm in the 
electromagnetic calorimeter is supported by the Russian Science Foundation 
(project No. 14-50-00080).

 \end{document}